\documentclass[10pt, conference, compsocconf]{IEEEtran}

\usepackage{cite}

\ifCLASSINFOpdf
\else
  \usepackage[dvips]{graphicx}
  \graphicspath{{../eps/}}
  \DeclareGraphicsExtensions{.eps}
\fi

\usepackage[caption=false]{caption}
\usepackage[font=footnotesize]{subfig}

\begin{document}
%
\title{Message transport characteristics in communication networks}


\author{\IEEEauthorblockN{Satyam Mukherjee}
\IEEEauthorblockA{Department of Physics\\
Indian Institute of Technology Madras\\
Chennai, India\\
Email: dirac.sat@gmail.com }
\and
\IEEEauthorblockN{Gautam Mukherjee}
\IEEEauthorblockA{Department of Physics\\
Asansol College\\
Asansol, India\\
Email: gautam.krishna@gmail.com}
\and
\IEEEauthorblockN{Neelima Gupte}
\IEEEauthorblockA{Department of Physics\\
Indian Institute of Technology Madras\\
Chennai, India\\
Email: gupte@physics.iitm.ac.in}
}


%


\maketitle

\begin{abstract}
We study message transport on a  $1-d$ ring  of nodes and randomly distributed hubs. Messages are deposited  on the network at a constant rate. When the rate at which messages are deposited on the lattice is very high, messages start accumulating after a critical time and the average load per node starts increasing. The power-spectrum of the load time-series shows $1/f$ like noise similar to the scenario of the Internet traffic. The inter-arrival time distribution of messages for the $1-d$ ring network shows stretched exponential behavior, which crosses over to power-law behavior if assortative connections are added to the hubs. The distribution of travel times in a related double ring geometry is shown to be bimodal with one peak corresponding to initial congestion and another peak to later decongestion.
\end{abstract}


\IEEEpeerreviewmaketitle

\section{Introduction}
Investigations of traffic flows on substrates of various topologies and discussions of  their efficiency have been a topic of recent research interest \cite{tadic,moreno}. The optimization of network structure and traffic protocols to achieve maximum efficiency is also a problem of practical importance. Congestion  effects can occur in real networks like telephone networks, computer networks and the Internet. Congestion/decongestion transitions are seen in these systems. Recent studies on the `ping'-experiment shows $1/f$ fluctuation at the critical point \cite {Takayasu}.

Message transport on different network geometries has been studied earlier on a linear chain \cite {Huisinga}, on two-dimensional lattices \cite{sat} and on Cayley trees \cite {Arenas} where messages are routed through shortest paths. Here,  we consider a $1-d$ ring lattice of ordinary nodes and hubs similar to that considered in \cite{sat}. Networks based on ring geometries have been studied in the context of ATM networks and Local Area Networks (LAN). A realistic ring topology such as fiber distributed data interface (FDDI) sends messages clockwise or counter-clockwise through the shared link. 
Similarly, messages are deposited  on our model  $1-d$ ring lattice at regular intervals. We show that the network reproduces the experimental findings of the Internet traffic flow.

\begin{figure*}[!t]
\centerline{\subfloat[]{\includegraphics[width=2.0in]{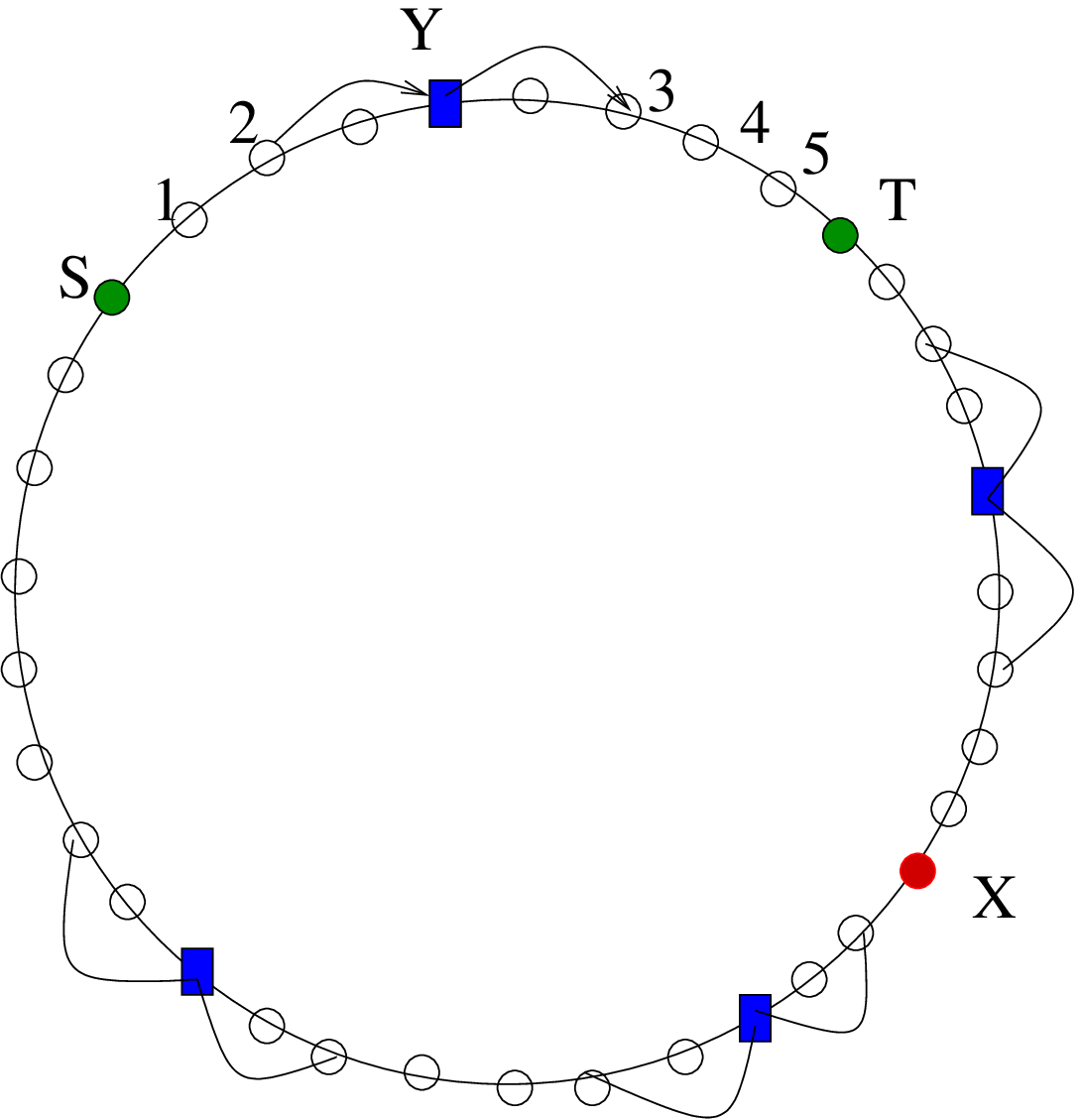}
\label{fig_1}}
\hfil
\subfloat[]{\includegraphics[width=3.0in]{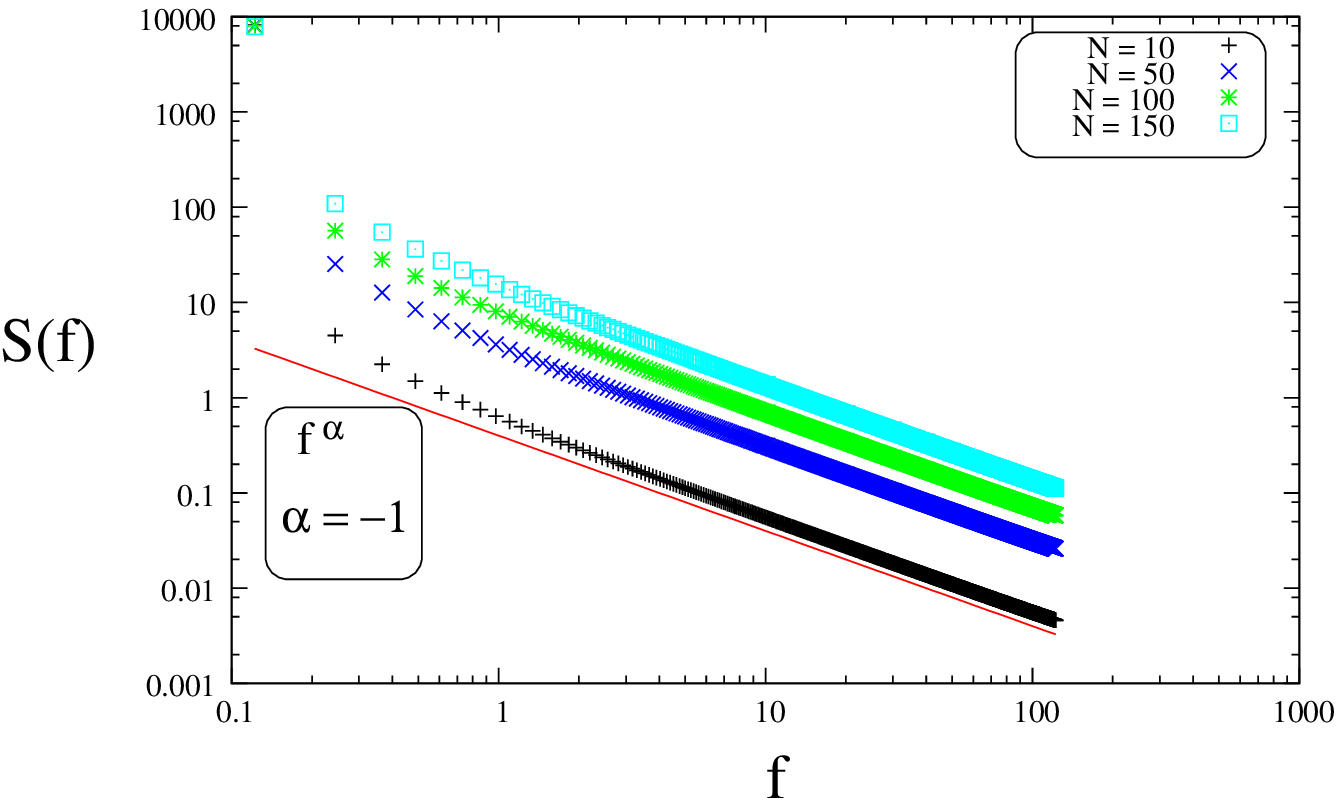}
\label{fig_2}}}
\caption{(a) A $1-d$ ring lattice of ordinary nodes ($X$) with nearest neighbor connections and randomly distributed hubs ($Y$). Each hub has $2k$ nearest neighbors, where $k=2$. (b) The plot of $S(f)$ against $f$ for posting rates of $N_{m}=10, 50, 100, 150$.}
\label{fig_sim1}
\end{figure*}

\section{A $1-d$ model of a communication Network} 

Here  we discuss an one dimensional version of the communication network of nodes and hubs. The base network is a ring lattice of size $L$ with nearest neighbor connections. Hubs are distributed randomly in the lattice where each hub has $2k$ nearest neighbors. No two hubs are separated by a less than a minimum distance, $d_{min}$. In our simulation we have taken $k$=4 and $d_{min}$=1, although Fig.\ref{fig_sim1}(a) illustrates only $k$=2 connections. The distance between a source and target is defined by the Manhattan distance $D_{st}=|is -it|$. Messages are routed along the shortest path between a source $S$ and a target $T$ in the clockwise direction taking advantage of all links in the direction of the target.
Thus, if a message is routed from a source $S$ to a target $T$ on this lattice through the baseline mechanism, it takes the path $S$-1-2-Y-3-4-5-$T$ as in Fig.\ref{fig_sim1}(a).

\section{Power Spectrum Analysis}

In our simulation,  a given number $N_{m}$ of source and target pairs start sending $N_{m}$ messages  at every $100-th$ time step for a total run time of $500000$ for a  lattice size $L=10000$, and $D_{st}=2000$. The average load per node is given as $\bar p(N_{m},t)=R(N_{m},t)/L$ where $R(N_{m},t)$ is the total number of messages flowing on the lattice. For smaller values of the posting rate, the value of $\bar p(N_{m},t)$ is very small and the system is in the decongested phase. As the posting rate of messages is increased, the system attains the congested regime.

The autocorrelation function of the average load per node ($\bar p(N_{m},t)$) is defined as \cite{gautam}:

\begin {equation}
C(N_{m},t) = \frac { \langle {\bar p(N_{m},t')\bar p(N_{m},t+t')} \rangle - \langle {\bar p(N_{m},t')} \rangle^2}
      {\langle {\bar p^2(N_{m},t')} \rangle - \langle {\bar p(N_{m},t')} \rangle^2}
\end {equation}
The Fourier transform of the  autocorrelation function $C(N_{m},t)$ is known as the spectral density or power spectrum $S(f)$, and is defined as
\begin {equation}
 S(N_{m},f) = \int^{\infty}_{-\infty} e^{-ift}C(N_{m},t)dt
\end {equation}

We plot $S(f)$ against $f$ for posting rates of $N_{m}=10, 50, 100, 150$. The plot of $S(f)$ against $f$  shows a power law: $S(f) \sim f^{-\alpha}$. In this case the spectral exponent $\alpha = 1$ thus indicating $1/f$ scaling irrespective of the posting rate (Fig.\ref{fig_sim1}(b)).

\begin{figure*}
\centerline{\subfloat[]{\includegraphics[width=2.0in]{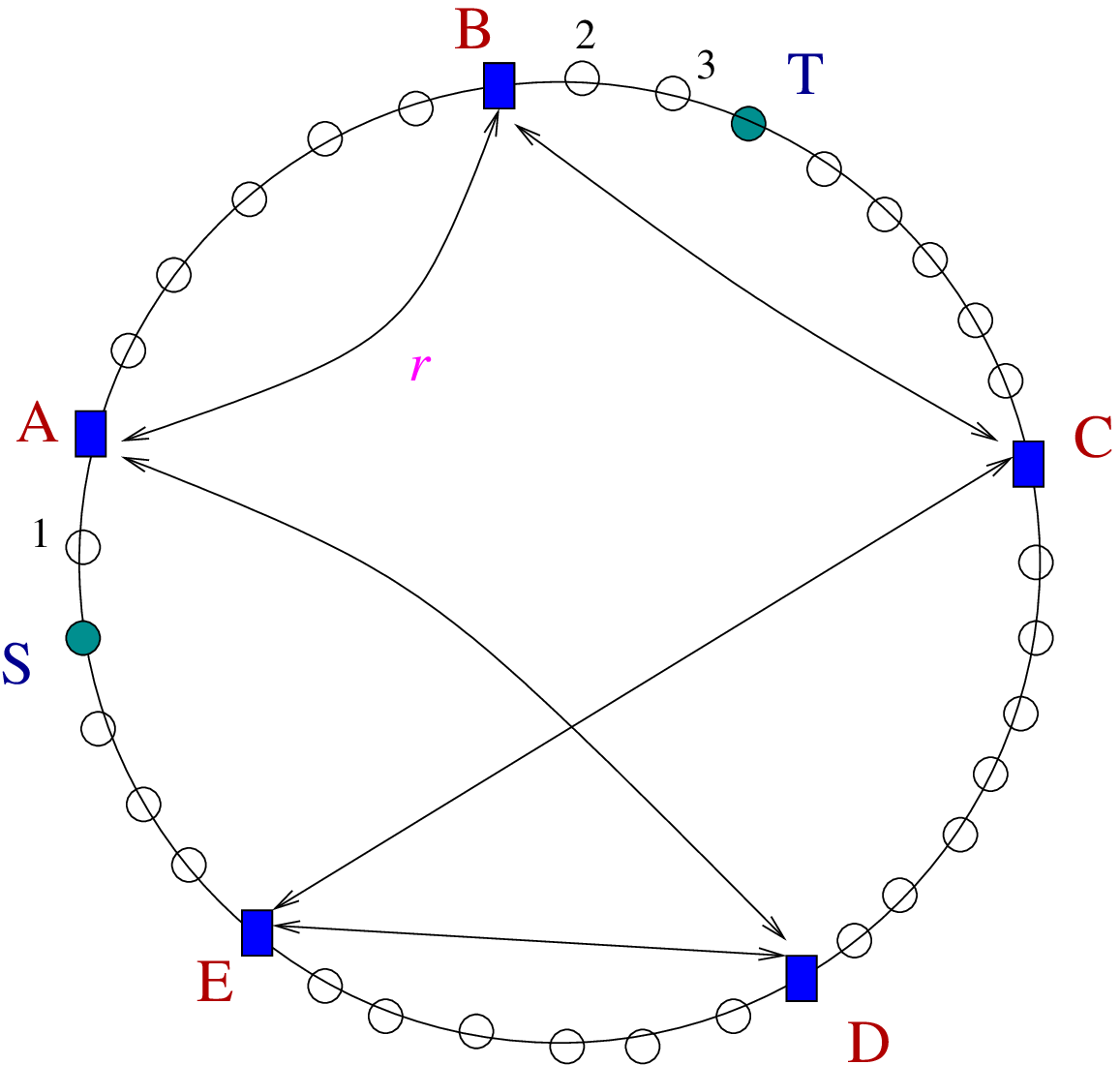}
\label{fig_5}}
\hfil
\subfloat[]{\includegraphics[width=3.0in]{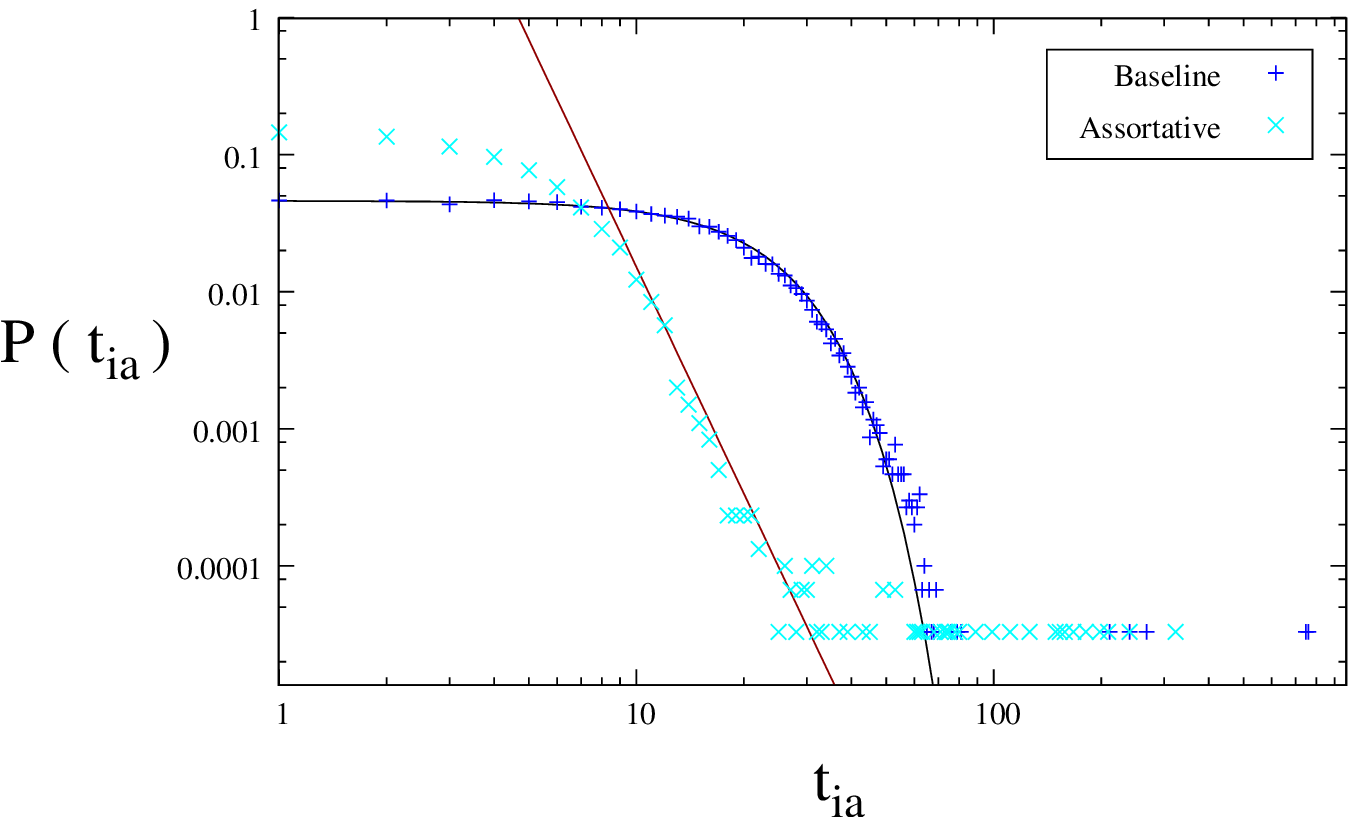}
\label{fig_6}}}
\caption{(a) The hubs are connected assortatively with two connections per hub. A message is routed along the path $S$-1-A-r-B-2-3-$T$. (b) The inter-arrival time distribution which shows a stretched exponential behavior for the baseline mechanism and a power law tail for the random assortative mechanism.} 
\label{fig_sim3}
\end{figure*}

We also study the inter-arrival time of messages for the most congested hub. The most congested hub is identified by calculating the coefficient of betweenness centrality (CBC), which is defined as the ratio of the number of messages $N_{k}$ which pass through a given hub $k$ to the total number of messages which run simultaneously i.e. $CBC=\frac{N_k}{R}$. Hubs with higher CBC value are more prone to congestion. We calculate the inter arrival time of messages for the hub with highest CBC. Inter-arrival times were studied earlier in the context of dynamics of information access on the web \cite{bara1} and also for human dynamics \cite{bara2}. For the baseline mechanism, the distribution of inter-arrival times is of
the stretched exponential form, given by

\begin{equation}
P(t_{ia})=\exp(-b{t_{ia}}^{\delta})
\end{equation}
where $\delta = 2.0$ for $N_{m}=50$ (Fig.\ref{fig_sim3}(b)). If the hubs in the lattice are connected by random assortative connections with two connections per hubs as shown in Fig.\ref{fig_sim3}(a), the inter arrival time of messages show power law behavior of the form
\begin{equation}
P(t_{ia})={t_{ia}}^{-\alpha}
\end{equation}
where $\alpha = 5.5$ for $N_{m}=50$ (Fig.\ref{fig_sim3}(b)). In the next section, we will discuss a double ring variation of the $1-d$ network, and discuss another statistical characteriser, the travel time distribution.

%

\section{The $1-d$ double ring communication network}

\begin{figure*}[!t]
\centerline{\subfloat[]{\includegraphics[width=2.0in]{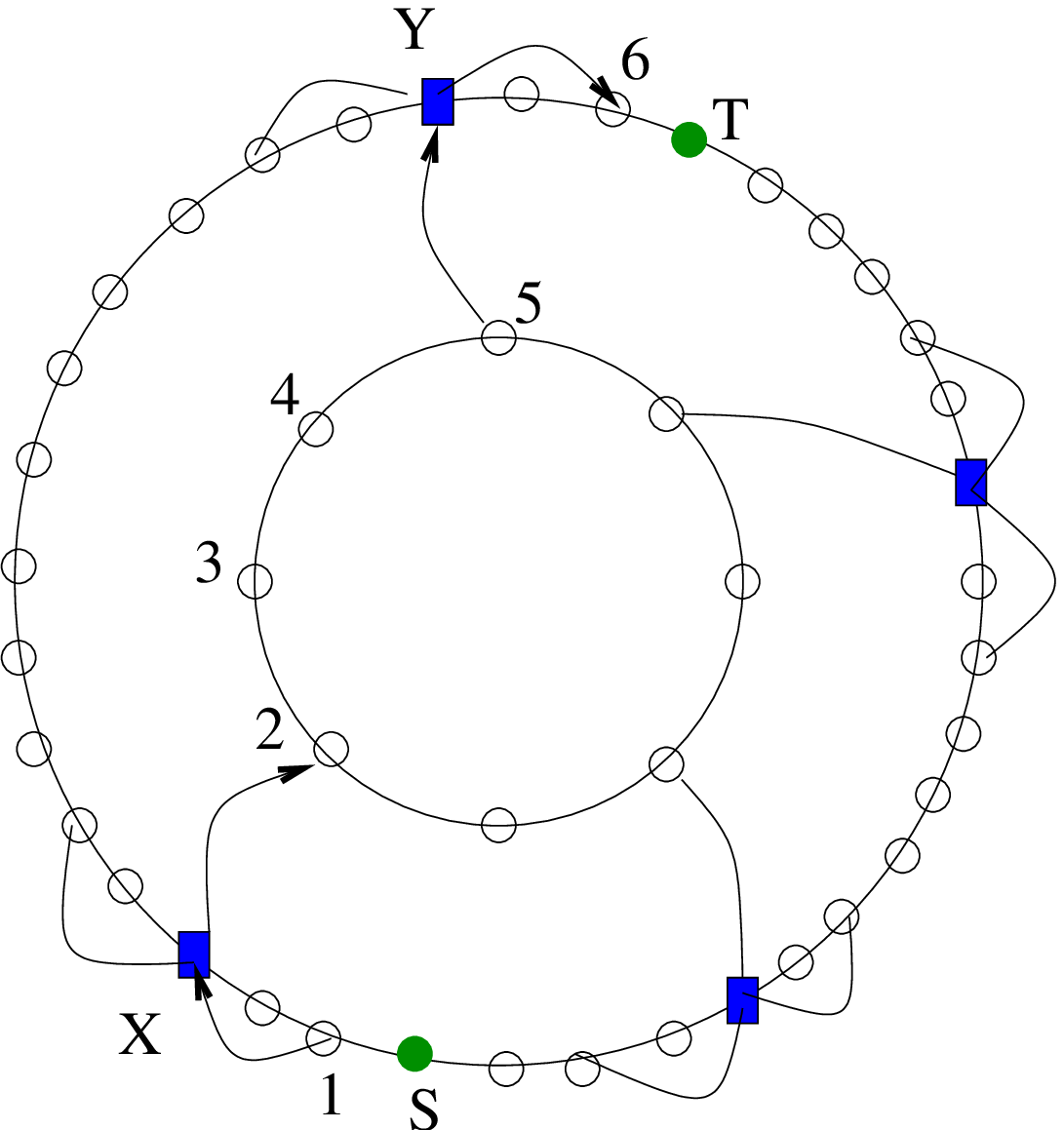}
\label{fig_3}}
\hfil
\subfloat[]{\includegraphics[width=3.0in]{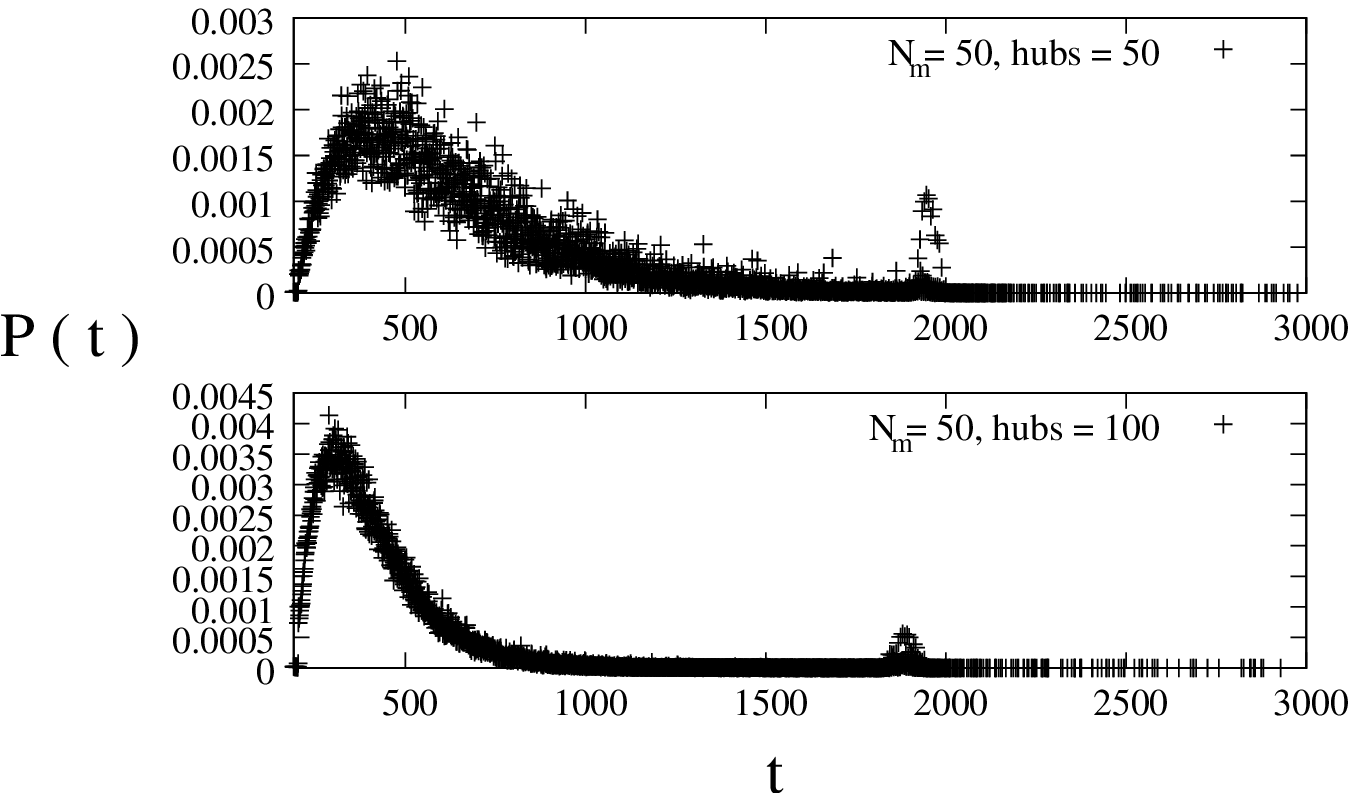}
\label{fig_4}}}
\caption{(a) Figure shows a $1-d$ double-ring lattice consisting of two concentric ring lattices. The outer ring consists of ordinary nodes with nearest neighbor connections and randomly distributed hubs ($X$, $Y$). Each hub has $2k$ nearest neighbors, where $k=2$. Each hub is connected to an ordinary node in the inner ring lattice. (b) The travel time distribution of messages flowing on this lattice shows bimodal distribution.}
\label{fig_sim2}
\end{figure*}

The $1-d$ ring lattice can be easily modified to the $1-d$ double-ring lattice as shown in Fig.\ref{fig_sim2}(a). Double-ring network topologies have been used earlier to model the head-direction system in animals \cite{seung} as well as for Local Area Networks (LAN). Our double-ring lattice consists of two concentric ring lattices (Fig.\ref{fig_sim2}(a)) of size $L_{i}$ and $L_{o}$ respectively, where $L_{i}$ is the size of the inner ring lattice and $L_{o}$ is the size of the outer ting lattice. The source-target pairs and the hubs are located in the outer lattice, with each hub having a connection to a node in the inner lattice. As before a message is routed along the shortest path $S$-1-X-2-3-4-5-Y-6-$T$ in the clockwise direction as shown in Fig.\ref{fig_sim2}(a). 

We study the travel time distribution of messages which are flowing on the lattice. The travel time is defined to be the time required for a message to travel from source to target, including the time spent waiting at congested hubs. A given number $N_{m}$ of source and target pairs start sending $N_{m}$ messages continuously at every $100$ time steps for a total run time of $30000$. In our simulation the travel time is calculated for a source-target separation of  $D_{st}=2000$ on a $L_{i}=1000$ and $L_{o}=9000$ double ring lattice, and averaged over $200$ hub realizations. The distribution of travel times of messages shows bimodal behaviour. The peak at higher travel times shows  Gaussian behavior whereas the peak at lower travel time shows log-normal behavior.  In the case of the $1-d$ ring, crossover from Gaussian to log-normal behavior was observed during the congestion-decongestion transition in the $1-d$ ring lattice \cite{sat2}. Hence we conclude that the Gaussian peak at higher travel times for the double ring corresponds to the initial congestion in the system, whereas the log-normal peak at lower travel times corresponds to the later decongested stage.

\section{Conclusion}
To summarize, we have studied message transport on model communication network of ordinary nodes and hubs, embedded on a ring lattice. The properties of message traffic on such a lattice is largely consistent with the real world networks like the Internet. The power spectral analysis of load time series data shows $1/f$ type fluctuations confirming long-ranged correlation in the network load time series, which is also seen in real life networks. For the baseline mechanism the inter arrival time distribution of messages show a stretched exponential behavior. The behavior changes to a power law if random assortative connections are introduced in the lattice. We also studied a variation of the ring lattice, namely the double ring lattice. The travel time distribution is bimodal, with one Gaussian peak and one log-normal peak. It would be interesting to see if our results have relevance in real life communication networks like telephone networks, biological networks etc.

\section*{Acknowledgment}
We  thank CSIR, India for support under their extra-mural scheme. The authors also thank A. Prabhakar for helpful suggestions and comments.



%

\end{document}